 \documentclass[final,3p,times,sort&compress]{elsarticle}

\usepackage{multirow,setspace,times,amssymb,amsmath,graphicx,color,rotating,subfigure,url}
\usepackage{lineno}
\usepackage{natbib}

\usepackage[labelfont=bf,labelsep=period,justification=raggedright]{caption}

\makeatletter
\renewcommand{\@biblabel}[1]{\quad#1.}
\makeatother
\date{}

\begin{document}
\begin{frontmatter}

\title{Evolution of citation networks with the hypergraph formalism}
\author[QH]{Feng Hu}
\author[QH]{Hai-Xing Zhao}
\author[NU]{Xiu-Xiu Zhan}
\author[ARC]{Chuang Liu}
\author[ARC]{Zi-Ke Zhang \corref{cor}}
\cortext[cor]{zhangzike@gmail.com}

\address[QH]{School of Computer, Qinghai Normal University, Xining 810008, China}
\address[NU]{Department of Mathematics, North University of China, Taiyuan 030051, China}
\address[ARC]{Alibaba Research Center for Complexity Sciences, Hangzhou Normal University, Hangzhou 311121, China}


\begin{abstract}

In this paper, we proposed an evolving model via the hypergraph to illustrate the evolution of the citation network. In the evolving model, we consider the mechanism combined with preferential attachment and the aging influence. Simulation results show that the proposed model can characterize the citation distribution of the real system very well. In addition, we give the analytical result of the citation distribution using the master equation. Detailed analysis showed that the time decay factor should be the origin of the same citation distribution between the proposed model and the empirical result. The proposed model might shed some lights in understanding the underlying laws governing the structure of real citation networks.
\end{abstract}

\end{frontmatter}



\section*{Introduction}

In recent years, more and more researchers focus on the study of the citation networks, which can depict the topological interactions between the academic publications \cite{Martin2013,Wangx2013} and the propagation of the scientific memes \cite{Kuhn2014}. Citation of scientific paper is a very significant gauge to measure the paper's importance \cite{Wang2013,Hirsch2005}. In order to understand the citation network structure more clearly, a series of evolving mechanism were presented. The first mathematical model of citation networks was proposed by Price in 1965 \cite{Price1965}, where the papers were depicted as the nodes and connections between nodes were depicted as directed edges in the citation networks. It noted that most papers received a small number of citations while very a few papers were cited many times. In addition, Price proposed the \emph{cumulative advantage process} \cite{Price1976} to illustrate the \emph{rich-get-richer} phenomenon in citation networks, which is also referred to as \emph{preferential attachment} mechanism \cite{Barabasi1999}. It is assumed that the papers with more citations were more likely to be cited again in the future, leading to the power-law like citation distribution ($p(k)\varpropto k^{-\gamma}$). And such preferential attachment mechanism also applied in the scientific collaboration networks to generate the power-law degree distribution of the collaborator numbers \cite{Newman2001}, where the probability of a particular scientist acquiring new collaborators increases with the number of his or her past collaborators. It is widely accepted that the preferential attachment is the basic mechanism to state the \emph{fat-tail} phenomenon in citation networks.

It should be noted that the power-law exponents in most empirical studies were smaller than that generated in the original BA model \cite{Jeong2003}. Furthermore, many citation networks derived from real world datasets do not follow a simple scaling solution, which exhibits that only the preferential attachment is not enough to describe the citation process. Many researches show that there is very strong relation between the node degree and its age. On the one hand, the citation process shows strong \textit{first move effect} \cite{Newman2009}. It means that the first published papers are expected to receive much more citations on average than those published later, because the first published papers are always the origin of the corresponding field. On the other hand, people are more likely to cite the new published papers which show the current research hotspot. This phenomenon generates that the connection probability of the new site with the old one is proportional not only to the connectivity of the old site but also to the power of its age, such as $\tau^{-\alpha}$ \cite{Dorogovtsev2002,Dorogovtsev2000}. In addition, Medo \emph{et al} \cite{Medo2011} proposed a growing network model combined preferential attachment and temporal eggects, which can be used for modeling a wide range of real systems. It is accepted that node age is a very important factor to affect the evolution of the citation networks, but how dose the node age influent the node degree is still confused.

Using the simple directed graphs to represent citation networks does not provide a complete description of the real systems \cite {Redner1998}. Recently, the hypergraph theory \cite{Berge1973}, which allows a hyperedge to connect an arbitrary number of vertices instead of two in regular graphs, has attracted much attention. It can provide us a promising way to understand real systems, such as the social networks \cite {Estrada2006}, reaction and metabolic networks \cite {Krishnamurthy2004}, protein networks \cite {Ramadan2004}, food webs \cite {Pimm1982}, social tagged networks \cite {Zlatic2009, Zhang2010}, scientific collaboration networks \cite {HuFeng2013,Chakraborty2013} and so on. Liu \emph{et al} \cite{Liu2014} proposed a knowledge-generation dynamic evolving model using the hyperedge growth and the hyperdegree preferential attachment mechanisms, which generates a power-law hyperdegree distribution in citation networks. In order to illustrate the citation process more clearly, we presented an evolving model combined the preferential attachment and the time decay via the hypernetwork. It is obtained that the hyperdegree distribution in the simulation result is consistent with the citation distribution in the American Physical Society ($APS$) dataset. Detailed analysis shows that the similar decay factor should be the reason for the same distribution between the simulation and the empirical result.

\section*{\label{Sec:Method}Method}

\subsection*{Models}

In the citation network based on the regular graph definition, a node represents a paper and a directed edge links two papers when one paper cited the other. The traditional evolving models are usually based on the BA model, where the new node always connects $m$ old nodes according to their connectivity in each step. It is easily to realize that the evolving model based on such representation has some limitations to describe the citation network. (i) It dose not consider the number of references that the new published paper cited. In general, the average number of references per paper increase significantly for a much easier way to read papers through the internet \cite{laki2014}. (ii) In the citation network, the new paper with no citation before won't be selected according to the BA model in the evolving process. In fact, people are more likely to cite the nearest papers in the real citation networks. In \cite{Dorogovtsev2000}, the citation network is treated as the undirect network to treat with the non citation papers. However, the large number of reference has nothing to do with the citation of the corresponding paper.

Therefore, we proposed an evolving model via hypergraph to tackle these limitations. We begin our study with some related definitions of hypergraph in the citation networks. Formally, a hypergraph $H$ can be depicted by $H=(V,E)$, where $V=\{v_1, v_2, ..., v_N\}$ is the set of nodes (or vertices), and $E=\{E_1, E_2, ..., E_e\}$  is the set of hyperedges, which contain an arbitrary number of nodes. And it is easily to realized that $E_i\neq\phi$, and $\bigcup_{i=1}^e=V$. A $k$-uniform hypergraph is a hypergraph that all hyperedges are formed with $k$ nodes. In this way, the $2$-uniform hypergraph is just the regular graph, a $3$-uniform hypergraph is a collection of triples \cite{Zhang2010}. Analogously to the regular networks, the hyperdegree of the node in the hypergraph is also the number of the hyperedges that contain the corresponding node.

\begin{figure}
  \centering
  \includegraphics[width=5cm]{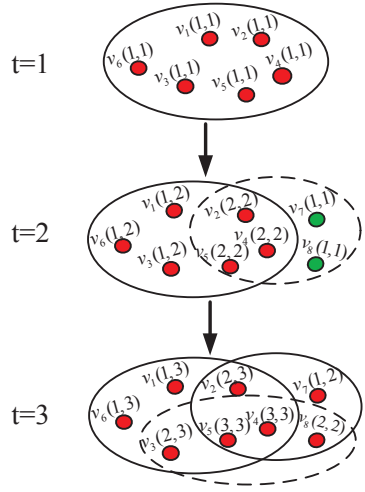}
  \caption{(Color online)\label{Fig:fmodel} Illustration of the hypergraph evolution. The ellipses and the filled circles represent the hyperedges and nodes respectively. $v_i(k,\tau)$ indicates that the hyperdegree of the node $v_i$ with age $\tau$ is $k$. The dash line and green nodes represent the new adding hyperedge and nodes respectively in each time step.}
\end{figure}

In the hypergraph formalism of the citation netowrk, the papers are considered as nodes and all the references of a paper can be described as a hyperedge. In this way, when a new paper is published, a hyperedge is added into the system. The citation number of a paper is just the node's hyperdegree in the hypergraph. It should be noted that the hyperedge include two types of references: one is the paper with citation number $\geq 1$ which is referred to as ``the old node", and the other is the paper that have not been cited before which is referred to as ``the new node". Considering the preferential attachment and the paper's influence decay in the citation networks, the probability that the new paper cites the existing one is related to not only the citation number but also its age. Thus we could construct the citation networks as follows (Fig. \ref{Fig:fmodel}):

\begin{itemize}
  \item There are $M_0$ nodes and a hyperedge including all these $M_0$ nodes in the system at the initial time (Fig. \ref{Fig:fmodel} (a)).
  \item A new paper with $L$ references is published at each time step and all the references construct a new hyperedge. In this way, a hyperedge with $L$ nodes enters into the network (Fig. \ref{Fig:fmodel} (b) and (c) ).
  \item Among the $L$ nodes, there are $m$ ($m\leqslant L$) ``old nodes" which determined by the probability $q_m$. The probability of selecting the old node $i$ (which entered into the network at time $t_i$) is:
        \begin{equation}\label{eq1}
        \centering
           \Pi(i,t)=\frac{k_i(t) \tau_i^{-\alpha}}{\sum\limits_{j} k_j(t) \tau_j^{-\alpha}},
        \end{equation}
        where, $\tau_i=t-t_i+1$ is the age of nodes $i$, $k_i(t)$ is the hyperdegree (citation number) of node $i$ at time $t$, and $\alpha$ is a tunable parameter which indicates the effect strength of the age. Generally speaking, the paper's influence would decay with time and people would be more likely to cite the new paper, leading to $\alpha>0$.
  \item The rest $L-m$ nodes in this hyperedge are ``new nodes" which haven't been cited before.
\end{itemize}

\subsection*{Analytical Analysis}

We analytically analysis the citation distribution with the evolving process aforementioned based on the master equation. In Eq. (\ref{eq1}), we defined $\Omega(t)=\sum\limits_{j} k_j(t) \tau_j^{-\alpha}$ which is the contribution of all nodes in the system, and it is the same for the each node in every time step. $\Omega(t)$ increased with time $t$, but as our consideration $\alpha>0$, we can obtain that $\lim\limits_{\tau\rightarrow\infty} \tau^{-\alpha} =0$, leading to the convergence value of $\Omega(t)$. And we can set $\Omega^*=\lim\limits_{t \rightarrow\infty}\Omega(t)$, which is a constant. Then we can obtain the master equation of hyperdegree distribution as follows:
\begin{equation}\label{eq4}
  p(k;i,\tau +1)=(1-M\frac{k \tau ^{-\alpha}}{\Omega ^*})p(k;i,\tau)+M \frac{(k-1)\tau ^{-\alpha}}{\Omega ^*}p(k-1;i,\tau),
\end{equation}
where $p(k;i,\tau)$ is denoted as the probability that node $i$ with age $\tau$ has hyperdegree $k$, and $M=\sum\limits_{m=1}^{L} mq_m$ is the expected number of a paper selected the ``old nodes". In order to depict the aging influence of the citation network, we assume that $m$ and $q_m$ is given, leading to the fixed $M$ in this process. In this equation, the first term is the probability of not selecting papers with $k$ citations and the second term is the probability of picking up papers with $k-1$ citations.

The hyperdegree distribution of nodes with age $\tau$ of the entire network is
\begin{equation}\label{eq5}
  p(k,\tau)=\sum \limits_{i\in V} p(k;i,\tau)/\tau,
\end{equation}
where, $V$ is the set of the nodes in the corresponding time.

Summing up eq.(\ref{eq4}) over $i$ through all nodes in the system, we get:
\begin{equation}\label{eq6}
  (\tau+1)p(k,\tau+1)=(1-rk\tau^{-\alpha})\tau p(k,\tau)+r(k-1)\tau ^{1-\alpha}p(k-1,\tau),
\end{equation}
where, $r=\frac{M}{\Omega^*}$ is also a constant.

We denote $p_k(\tau)=p(k,\tau)$ when $\tau\rightarrow\infty$. And at long times, we can obtain the following differential equation:
\begin{equation}\label{eq7}
  \tau\frac{dp_k(\tau)}{d\tau}+(1+rk\tau^{1-\alpha})p_k(\tau)=r(k-1)\tau ^{1-\alpha}p_{k-1}(\tau),
\end{equation}
with the boundary conditions that $p_k(1)=1$ for $k=1$, whereas for $k>1$, $p_k(1)=0$, and $p_k(0)=0$.

Following the method in Ref. \cite{Newman2009}, we can obtain the solution of Eq. (\ref{eq7}) as follows:
\begin{equation}\label{eq12}
  p_k(\tau)=\frac{1}{\tau}exp((1-\tau^{(1-\alpha)})\frac{r}{1-\alpha})(1-exp((1-\tau^{(1-\alpha)})\frac{r}{1-\alpha}))^{k-1},
\end{equation}
Equation (\ref{eq12}) gives us the general solution for the probability distribution of a paper's citations at age $\tau$. The overall distribution of citations over the age from 1 to $\tau_0$, which we denoted as $P_k(\tau_0)$, can be calculated as:

\begin{equation}\label{eq13}
\begin{split}
  P_k(\tau_0)&=\frac{1}{\tau_0}\int\limits_{1}^{\tau_0} p_k(\tau)d\tau  \\
  &=\frac{1}{\tau_0}\int\limits_{1}^{\tau_0} \frac{1}{\tau}exp((1-\tau^{(1-\alpha)})\frac{r}{1-\alpha})(1-exp((1-\tau^{(1-\alpha)})\frac{r}{1-\alpha}))^{k-1} d\tau
\end{split}
\end{equation}

We assumed $u=exp((1-\tau^{1-\alpha})\frac{r}{1-\alpha})$, and the Eq. (\ref{eq13}) can be written as follows:
\begin{equation}\label{eq15}
  P_k(\tau_0)=-\frac{1}{\tau_0}\frac{1}{r}\int\limits_1^{u_0} (1-\frac{1-\alpha}{r}lnu)^{-1}(1-u)^{k-1}du,
\end{equation}

Using the Taylor expansion $1-\frac{1-\alpha}{r}lnu\approx u^{-\frac{1-\alpha}{r}}$, we can get:
\begin{equation}\label{eq16}
  P_k(\tau_0)\approx -\frac{1}{\tau_0}\frac{1}{r}\int\limits_1^{u_0} u^{\frac{1-\alpha}{r}}(1-u)^{k-1}du,
\end{equation}
where $u_0=exp((1-\tau_0^{1-\alpha})\frac{r}{1-\alpha})$.

With the substitution $q=1-u$, the Eq. (\ref{eq16}) can be rewritten as follows:
\begin{equation}\label{eq17}
  P_k(\tau_0)\approx \frac{1}{\tau_0}\frac{1}{r}\int\limits_0^{1-u_0} q^{k-1}(1-q)^{\frac{1-\alpha}{r}}dq,
\end{equation}

$P_k(\tau_0)$ is the regularized incomplete bata function \cite{Gautschi1967} and $u_0=exp((1-\tau_0^{1-\alpha})\frac{r}{1-\alpha})$, hence
\begin{equation}\label{eq18}
  P_k(\tau_0)\approx A\frac{1}{k}(1-exp((1-\tau_0^{1-\alpha})\frac{r}{1-\alpha}))^k,
\end{equation}



\begin{figure}
  \centering
  \includegraphics[width=10cm]{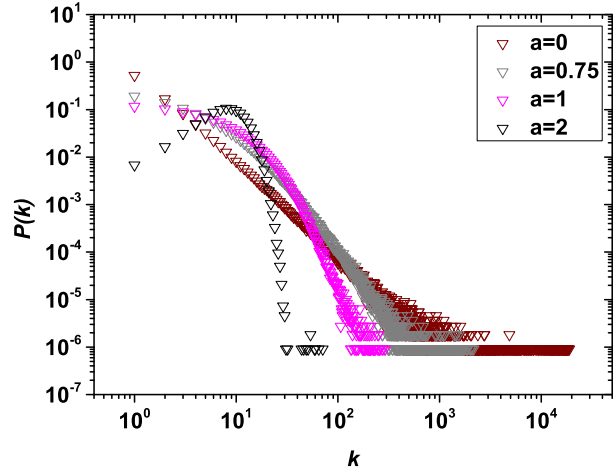}
  \caption{(Color online)\label{Fig:hfig1} The hyperdegree distribution for different values of $\alpha$. 1.)$\alpha=0$; 2.)$\alpha=0.75$; 3.)$\alpha=1$; 4.)$\alpha=2$; }
\end{figure}

\section*{Results}

We are mainly interested in the influence of the age in the evolving model. In this way, apart from the parameter $\alpha$, the other parameters are set as follows: $L=10$, and $m=(6,7,8,9,10)$ which is corresponding to $q_m=(0.05,0.05,0.05,0.15,0.70)$ respectively. And the expect value of the number of ``old nodes" is $M=\sum\limits_{m=1}^{L} mq_m=9.4$, which means that there are about $0.6$ ``new nodes" are cited each step in average. It is reasonable that the mean value of the cited ``new nodes" is smaller than 1 for we add just one paper each step. Here, we consider the region $\alpha\geq 0$, because only the region $\alpha \geq 0$ seems to be of real significance. Fig. \ref {Fig:hfig1} demonstrates the hyperdegree distribution with different $\alpha$ values. When $\alpha=0$, we can find that the hyperdegree distribution shows a power-law dependence $P(k)\propto k^{-\gamma}$ (as the red inverted triangle in Fig. \ref{Fig:hfig1}). It coincides with the model description as Eq. (\ref{eq1}), which is just the BA model when $\alpha=0$. The result shows that the aging factor influences the hyperdegree distribution significantly, and the scaling of the hyperdegree distribution disappeared when $\alpha>0$. The preferential selecting properties in the evolving process vanished when $\alpha$ is large enough, leading to the moderate hyperdegree distribution (as the pink inverted triangle in Fig. \ref{Fig:hfig1}).

\begin{figure}
  \centering
  \includegraphics[width=10cm]{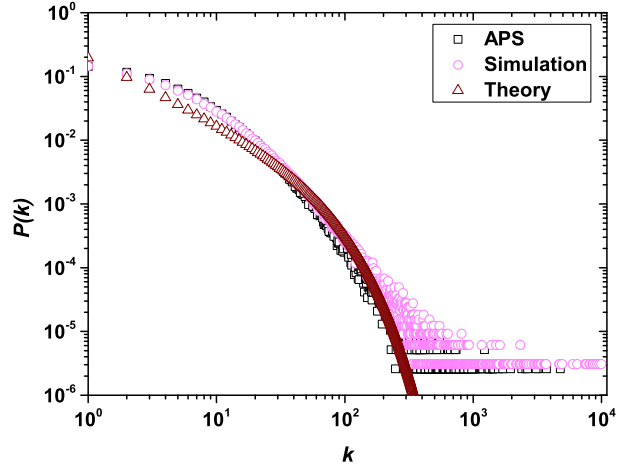}
  \caption{(Color online)\label{Fig:hfig2} The hyperdegree distribution of nodes of three cases: APS data (black square), simulation (pink circle) and theory analysis (red triangle). The simulation result is obtained with $\alpha=0.75$.}
\end{figure}

In order to test the proposed model, the American Physical Society ($APS$) data set is adopted. There are $463442$ papers published from $1893$ to $2009$ in the data. The total citation is $4708753$, and each paper is cited $10.2$ in average \footnote{Details about the data set can be found on the web at https://publish.aps.org/datasets}. We show the citation distribution of the APS data (blue square) in Fig. \ref{Fig:hfig2}. The simulation result (red circle) is agreement with the empirical data very well with the parameter $\alpha\approx 0.75$, which indicates that the evolving model with aging in the hypergraph formalism is reasonable. The positive $\alpha$ shows the accurate assumption coming from the common sense that people are likely to cite the resent papers. In addition, we also plot the theory result in Fig. \ref{Fig:hfig2}.

\begin{figure}
  \centering
  \includegraphics[width=14.5cm,height=8cm]{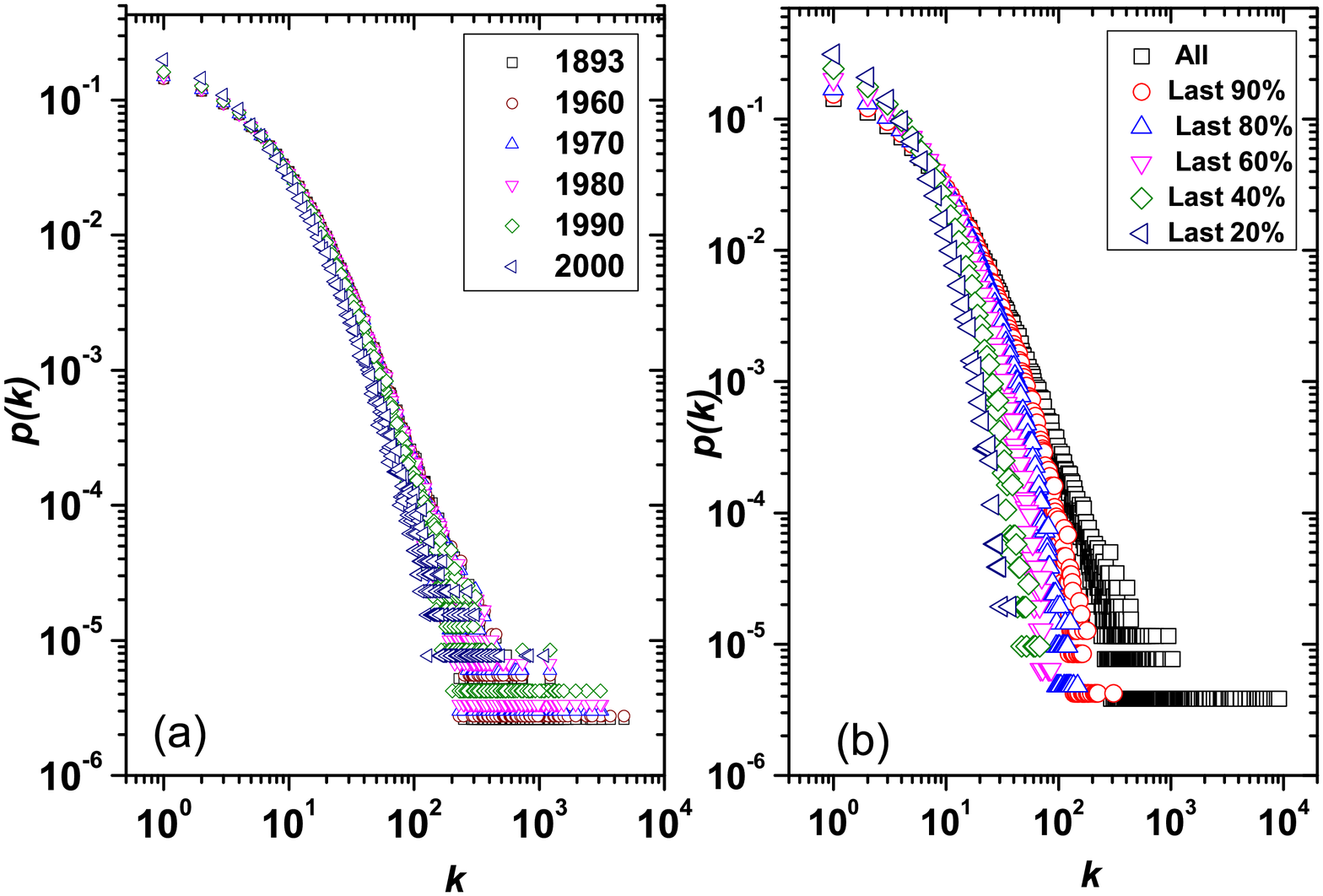}
  \caption{(Color online)\label{Fig:hfig4} Citation distribution of nodes in different age groups for the two cases: (a) $APS$ and (b)$simulation$.}
\end{figure}

In order to analyse the evolving process in detail, we study the hyperdegree distribution of the nodes in different age groups in Fig. \ref{Fig:hfig4}. For the APS data (Fig. \ref{Fig:hfig4} (a)), we consider the citation distribution of the papers that published after the corresponding year. For the simulation with parameter $\alpha=0.75$ (Fig. \ref{Fig:hfig4} (b)), we consider the hyperdegree distribution of the nodes in different age groups. In general, the most highly cited papers are always the earliest papers in this field, because of not only the first-move effect, but also the long term to accumulate citations. And the tail of the hyperdegree distribution exhibits the power-law form when we consider the oldest papers. In the converse, the recent papers usually possess few citations. When we exclude the earliest papers in the data set, it is tended to throw out the papers with large number of citations, and the tail of hyperdegree distribution changed from the power-law to exponential, which is consistent with the findings in Ref. \cite{Newman2009}.

\begin{figure}
  \centering
  \includegraphics[width=14.5cm,height=8cm]{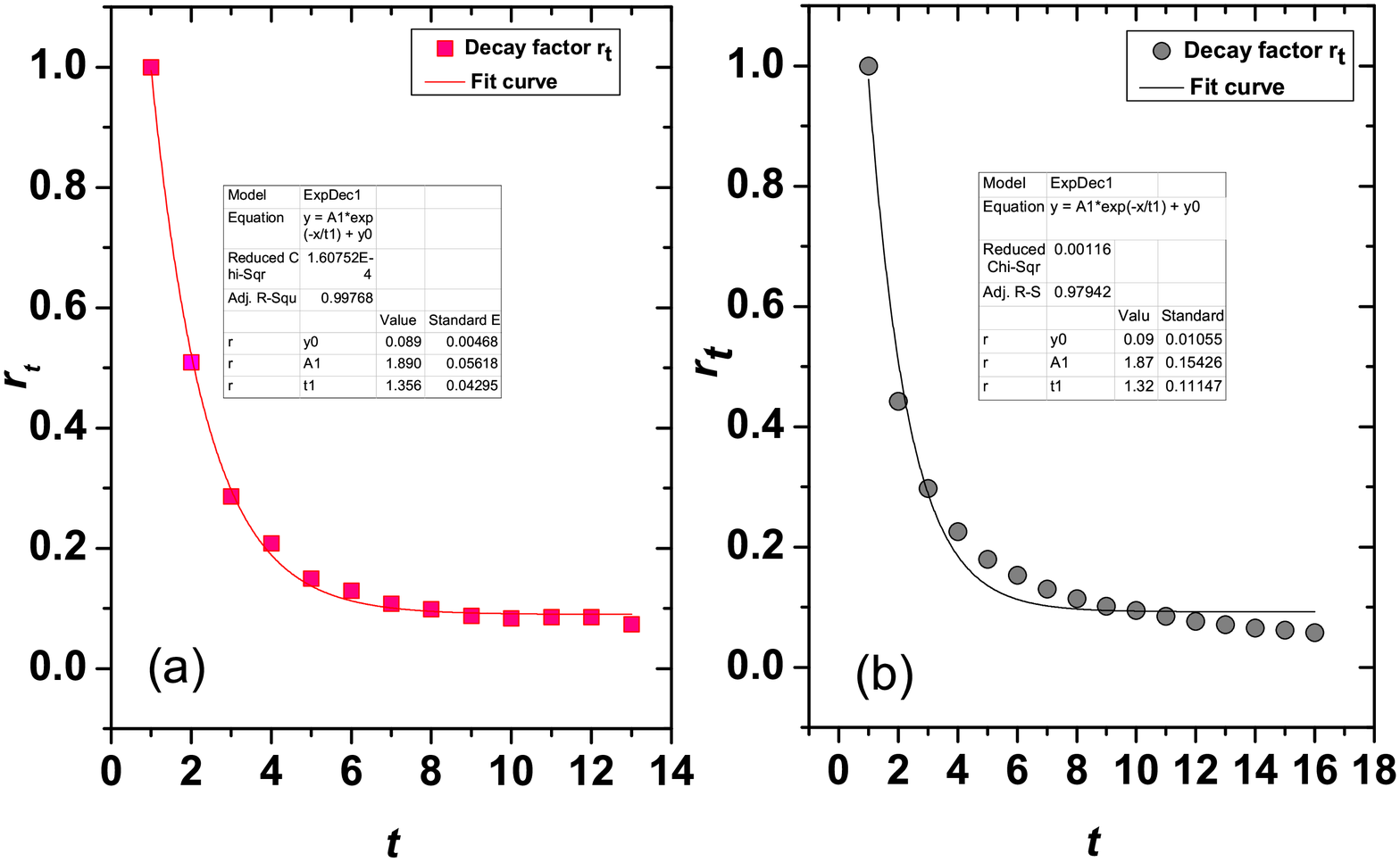}
  \caption{(Color online)\label{Fig:hfig5} The decay factor $r_t$ as a function of time. Solid curve is fit value.(a) $APS$ and (b)$simulation$.}
\end{figure}

It is interested to find that the model result fits the APS data very well when the parameter $\alpha=0.75$. As illustrated in the model description, the parameter $\alpha$ controls the decay rate of the paper's influence. Large $\alpha$ shows that the paper's influence decays very quickly, and vice verse. We guess that the time decay factor is also round $0.75$ in the APS data, which generates the similar hyperdegree distribution between the empirical and the model result. In order to verify the assumption, we analysis the decay factor $r_t$ value, which can be calculated as follows \cite{WuF2007}:
\begin{equation}\label{eq19}
  r_t=\frac{E(logN_t)-E(logN_{t-1})}{E(logN_1)-E(logN_0)}
\end{equation}
where $N_t$ is the citations of the considering papers at time $t$, and $E(\cdot)$ is the expected value of $(\cdot)$. Figure \ref{Fig:hfig5} displays the decay factor $r_t$ as a function of time $t$. The left panel shows the decay factor of the APS data, where we consider the citation increment of the papers that published before $1970$. We set two yeas as the time interval, and $N_t$ is the citations of these papers in each period of time. Analogously, the right panel indicates the decay factor of the evolution model, where we consider the hyperdegree increment of the nodes that added into the system at the initial 10000 time steps. We set 10000 steps as the time interval in this analysis, and $N_t$ is the citations of the corresponding papers at each period of time step. It is interesting to find that the APS data and the model result share the same decay factor function $r_t\propto e^{-\eta t}$ with $\eta=0.75$. The exponential decay of $r_t$ indicates that the paper's influence decrease very quickly, and it is consistent with that the quick updating of the scientific research achievement. The similar decay rate of $r_t$ should be the reason for the same hyperdegree distribution between the APS data and the model result with $\alpha=0.75$.

\section*{Conclusion and Discussion}

The preferential attachment is a basic hypothesis in the evolving network analysis. Recently, the aging of the nodes attract much more attention, because the evolution is essentially a dynamic process, and  in general the node attribute changes with the time. In this paper, we proposed an evolving model via the hypernetwork to illustrate the evolution of the citation network. In the evolving model, we consider the mechanism combined with preferential attachment and the aging influence. Simulation results show that the proposed model can characterize the citation distribution of the real network very well. In addition, we give the analytical result of the citation distribution using the master equation. Detailed analysis showed that the time decay factor should be the origin of the same citation distribution between the proposed model and the empirical result. The proposed model might shed some lights in understanding the underlying laws governing the structure of real citation networks.

\section*{Acknowledgements}
This work was partially supported by the Fund for Less Developed Regions of the National Natural Science Foundation of China (Grant No. 61164005), the National Natural Science Foundation of China (Grant Nos. 11105024, 11205040, 11305043, 1147015), the Nature Science Foundation from Qinghai Province (No. 2012-ZR-3061), the start-up research funding and Pandeng Project of Hangzhou Normal University.

\end{document}